\documentclass[reprint, amsmath, amssymb,aps,prl,floatfix,longbibliography]{revtex4-1}

\usepackage{amsmath, graphicx,natbib, bm ,array, bibentry, hyperref}
\hypersetup{colorlinks = true,allcolors = blue }

\allowdisplaybreaks

\begin{document}

\newcommand{\pd}{\partial}
\newcommand{\beq}{\begin{equation}}
\newcommand{\eeq}{\end{equation}}
\newcommand{\bseq}{\begin{subequations}}
\newcommand{\eseq}{\end{subequations}}

\newcommand{\coeffI}{Z_I}
\newcommand{\coeffP}{Z_P}
\newcommand{\fermifn}{f}

\title{Many-body wavefunctions for quantum impurities out of equilibrium}

\author{Adrian B. Culver}
\email[Present address: Mani L. Bhaumik Institute for Theoretical Physics, Department of Physics and Astronomy, University of California, Los Angeles, California 90095, USA; ]{adrianculver@physics.ucla.edu}
\author{Natan Andrei}
\email{natan@physics.rutgers.edu}
\affiliation{Center for Materials Theory, 
Department of Physics and Astronomy, Rutgers University, Piscataway, New Jersey 08854
}

\date{\today}

\begin{abstract}
We present a method for calculating the time-dependent many-body wavefunction that follows a local quench.  We apply the method to the voltage-driven nonequilibrium Kondo model to find the exact time-evolving wavefunction following a quench where the dot is suddenly attached to the leads at $t=0$.  The method, which does not use Bethe ansatz, also works in other quantum impurity models (we include results for the interacting resonant level and the Anderson impurity model) and may be of wider applicability.  In the particular case of the Kondo model, we show that the long-time limit (with the system size taken to infinity first) of the time-evolving wavefunction is a current-carrying nonequilibrium steady state that satisfies the Lippmann-Schwinger equation.  We show that the electric current in the time-evolving wavefunction is given by a series expression that can be expanded either in weak coupling or in strong coupling, converging to all orders in the steady-state limit in either case.  The series agrees to leading order with known results in the well-studied regime of weak antiferromagnetic coupling and also reveals another universal regime of \emph{strong ferromagnetic} coupling, with Kondo temperature $T_K^{(F)} = D e^{-\frac{3\pi^2}{8} \rho |J|}$ ($J<0$, $\rho|J|\to\infty$).  In this regime, the differential conductance $dI/dV$ reaches the unitarity limit $2e^2/h$ asymptotically at \emph{large} voltage or temperature.
\end{abstract}

\maketitle

\emph{Introduction.} The Kondo model,  which describes a localized spin interacting via spin exchange with one or more reservoirs of electrons, has long been a source of new ideas in theoretical physics \cite{Hewson}.  Its nonequilibrium physics became of great interest when the model was realized in quantum dot systems in the Coulomb blockade regime, with attached leads serving as reservoirs.  When a voltage difference is imposed between two leads, an electric current is driven through the dot \cite{Goldhaber-Gordon,Cronenwett,vanderWiel,KretininEtAl}.  The corresponding nonequilibrium theory has been studied by a variety of approaches, both in the Kondo model itself \cite{KaminskiNazarovGlazman, RoschEtAlPRL, DoyonAndrei, Schoeller, PletyukhovSchoeller, AshidaEtAl} and in the more general Anderson impurity model \cite{HershfieldDaviesWilkins, Oguri, KonikSaleurLudwig, WernerOkaMillis, Heidrich-MeisnerEtAl, EckelEtAl, ChaoPalacios, DordaEtAl, ProfumoEtAl, SchwarzEtAl} (see the references of Ref. \cite{AshidaEtAl} for a more complete list).

\begin{figure}[htp]
    \includegraphics[scale=.4]{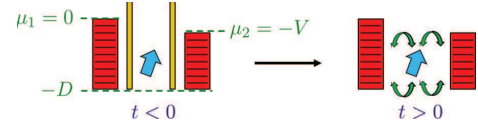}
    \caption{Schematic of the quench process.  Prior to $t=0$, the leads are filled with free electrons, with no tunneling to the dot allowed.  From $t=0$ onward, the system evolves with the many-body Hamiltonian $H$, with tunneling to and from the leads resulting in an electric current.}
    \label{fig: Quench}
\end{figure}

In this Letter, we probe the nonequilibrium physics via a quantum quench, a protocol in which the ground state of an initial Hamiltonian $H_i$ is evolved in time by a final Hamiltonian $H_f$ following a sudden change of parameters.   Here, the initial state of the quench consists of a free Fermi sea in each of the two leads---with the applied voltage appearing as the difference in chemical potentials---and the quench protocol  consists of evolving this state by the full Kondo Hamiltonian $H_f\equiv H$ (see Fig. \ref{fig: Quench}). This quench allows us to access the steady-state regime of the nonequilibrium Kondo model in the long-time limit (we first take the limit of infinite system size so that the leads serve as thermal baths \cite{DoyonAndrei}).

We present here the results of a method for calculating the many-body wavefunction following this quench. This method is also applicable to other quantum impurity problems and may have wider applicability.  We concentrate on the steady-state current as a function of the source-drain voltage and the temperatures of the leads.   We first calculate the current in the much-studied weak coupling antiferromagnetic regime and find, as expected, that at energy scales much smaller than the bandwidth, the current is a universal function governed by an emergent scale: the Kondo temperature $T_K$. 
We then proceed to identify another universal regime: \emph{strong ferromagnetic} coupling, with its own scale $T^{(\text{F})}_K$.  Further details on our calculations are available in Refs. \cite{CulverAndrei_PRB1,CulverAndrei_PRB2, Culver_thesis}.

With universality in mind, we study the two-lead Kondo model in the wide-band limit \cite{DoyonAndrei}:
\begin{multline}
    H = -i \int_{-L/2}^{L/2}dx\ \sum_{\gamma=1,2} \psi_{\gamma a}^\dagger(x) \frac{d}{d x}\psi_{\gamma a}(x)\\
    + \sum_{\gamma,\gamma'=1,2} \frac{1}{2}J \psi_{\gamma a}^\dagger(0)\bm{\sigma}_{aa'}\psi_{\gamma' a'}(0)\cdot \mathbf{S}.\label{eq: two-lead Kondo}
\end{multline}
Formally, $\rho = \rho_1 \otimes \rho_2$ is the initial density matrix, where  $\rho_\gamma = \exp \left[ -\frac{1}{T_\gamma}\sum_{|k|<D}(k-\mu_\gamma) c_{\gamma ka}^\dagger c_{\gamma ka} \right] $ is the Fermi distribution  (cut off by the bandwidth $D$) in lead $\gamma=1,2$, and $\rho(t) = e^{-i H t} \rho e^{i H t}$ is the time-evolving density matrix following the quench at $t=0$.

Our method provides the explicit and exact solution for $\rho(t)$.  The solution applies for $ 0\le t < L/2$, which is the regime of interest: In the calculation of the current, we take the steady-state limit ($t\to\infty$) \emph{after} the thermodynamic limit ($L\to\infty$ with $D$ fixed, hence a fixed density of electrons).  The thermodynamic limit is taken order by order either in $J$ or $1/J$.

\emph{Exact wavefunction.} It suffices to find the time evolution of an $N$-electron state (rather than density matrix) with arbitrary quantum numbers, $|\Psi(t)\rangle \equiv e^{-i H t} c_{\gamma_N k_N a_N}^\dagger \dots c_{\gamma_1 k_1 a_1}^\dagger|a_0\rangle$, where $a_0$ is the impurity spin.  Our method yields \footnote{The $n=0$ term of the sum is understood to be $\left( \prod_{j=1}^N e^{-i k_j t} c_{\gamma_j k_j a_j}^\dagger \right)|a_0\rangle$.}
\begin{multline}
    |\Psi(t)\rangle = \sum_{n=0}^N 2^{-n/2} \sum_{1\le m_1< \dots  < m_n \le N} (-1)^{ m_1+\dots+m_n +1 }\\
    \times \left( \prod_{j=1, j \ne m_{\ell}\ \forall \ell}^N e^{-i k_j t} c_{\gamma_j k_j a_j}^\dagger \right)  
    \sum_{\sigma \in \text{Sym}(n) } \left( \text{sgn}\ \sigma\right)\\
    \times |\chi_{k_{m_{\sigma(1)}} a_{m_{\sigma(1)}}\dots k_{m_{\sigma(n)}}a_{m_{\sigma(n)}} ,a_0}(t)\rangle,
\end{multline}
where $c_{\gamma k a}^\dagger= \int_{-L/2}^{L/2} dx\ \frac{e^{i k x}\psi_{\gamma a}^\dagger(x)}{\sqrt{L}}$,
\begin{multline}
    |\chi_{k_1 a_1 \dots k_n a_n,a_0}(t)\rangle = \int_0^t dx_1 \int_0^{x_1}dx_2\dots \int_0^{x_{n-1}}dx_n\
    \delta_{a_0}^{c_0} \delta_{c_n}^{b_0}\\
    \times \left[  \prod_{j=1}^n \frac{e^{-i k_j(t- x_j)}}{\sqrt{L}} 
    \times \left( -i \mathcal{T}_{a_j c_{j-1}}^{b_j c_j} \right)
    \psi_{e b_j}^\dagger(x_j) \right] |b_0\rangle,
\end{multline}
$\psi_{ea}^\dagger = \frac{ \psi_{1a}^\dagger + \psi_{2a}^\dagger}{\sqrt{2}}$,
and $\mathcal{T}_{a_1 a_0}^{b_1 b_0} = \biggr[ -\frac{1}{2}J \left(1 + i \frac{3}{4} J \right) \delta_{a_1}^{b_1}\delta_{a_0}^{b_0} +J  \delta_{a_1}^{b_0}\delta_{a_0}^{b_1}  \biggr] /( 1 - i\frac{1}{2}J + \frac{3}{16}J^2)$ (the $\mathcal{T}$ matrix for a single electron crossing the impurity).

With minor modifications, the solution can be extended to a more general model with an anisotropic Kondo interaction, a potential scattering term, and a magnetic field on the impurity \cite{CulverAndrei_PRB1, Culver_thesis}.

In the long-time limit (with the system size always larger), $|\Psi(t)\rangle$ becomes a nonequilibrium steady state (NESS): an energy eigenstate of $H$ with the boundary condition of incoming plane waves with the quantum numbers $\gamma_j k_j a_j$ (i.e., a many-body Lippmann-Schwinger ``in'' state).  The NESS can be found directly using a time-independent version of our method.

\emph{The current.} Using our exact answer for $\rho(t)$, we proceed to calculate the average electric current, $I(t) = -\frac{d}{dt} \text{Tr}\left[ \rho(t) \widehat{N}_1 \right]/\text{Tr}\rho$, where $\widehat{N}_1= \int_{-L/2}^{L/2}dx\ \psi_{1a}^\dagger(x) \psi_{1a}(x)$.  For a fixed system size $L$ and bandwidth $D$, we can write the current as a finite sum; however, taking the thermodynamic limit ($L\to\infty$) is a formidable task.  We take the limit order by order in an expansion parameter which can be either $J$ or $1/J$.  In this way, we arrive at a series answer that probes both the usual weak coupling regime of the model and a new strongly coupled regime.

In the thermodynamic limit, sums over momenta become integrals involving the Fermi functions  $\fermifn_\gamma(k)\equiv 1/(e^{(k-\mu_\gamma)/T_\gamma} +1)$ of the leads, resulting in a series expression for the current:
\begin{widetext}
    \begin{align}
        I(T_1, \mu_1;T_2,\mu_2; t) &= \text{Re} \Biggr\{ \frac{\partial}{\partial t} \sum_{n=1}^\infty \sum_{\sigma\in \text{Sym}(n) } W_n^{(\sigma)}(J)  \int_{-D}^D \frac{dk_1\dots dk_n}{(2\pi)^n} \int_0^t dx_1 \int_0^{x_1}dx_2\dots \int_0^{x_{n-1}}dx_n\ \notag \\
        &\qquad \qquad \qquad \times \left[ \prod_{j=1}^{n-1} \left[ \fermifn_1(k_j) + \fermifn_2(k_j) \right] e^{i ( k_{\sigma_j} - k_j)x_j} \right] \left[ \fermifn_1(k_n) - \fermifn_2(k_n) \right] e^{i ( k_{\sigma_n}- k_n )x_n}
        \Biggr\},\label{eq: final answer for current} \\
        \text{where } W_n^{(\sigma)}(J) &= \left( \text{sgn}\ \sigma\right) \frac{i}{2^{n+1}} \sum_{\substack{a_1,\dots,a_n \\ b_1, \dots, b_{n-1} \\ c_0,c_0' \dots, c_{n-1}, c_{n-1}'}} \delta_{c_0}^{c_0'} \mathcal{T}_{a_{\sigma_n} c_{n-1}}^{a_n c_{n-1}'}\prod_{j=1}^{n-1} \left( \mathcal{S}_{a_j c_{j-1}'}^{* b_j c_j'} \mathcal{S}_{a_{\sigma_j} c_{j-1}}^{b_j c_j} -\delta_{a_j}^{ b_j }\delta_{c_{j-1}'}^{c_j'} \delta_{a_{\sigma_j}}^{ b_j} \delta_{c_{j-1}}^{c_j} \right),
    \end{align}
\end{widetext}
with $ \mathcal{S}_{a_1 a_0}^{b_1 b_0} \equiv \delta_{a_1}^{b_1}\delta_{a_0}^{b_0} - i \mathcal{T}_{a_1 a_0}^{b_1 b_0}$ \footnote{Although no Bethe ansatz technology was applied, we see here the same bare $\mathcal{S}$ matrix that appears in the Bethe ansatz solution of the one-lead model in equilibrium (see Ref. \cite{Andrei_lecture_notes}, for example, bearing in mind that our convention is related by $J_{\text{Bethe ansatz}} = \frac{1}{2}J$).}.
It can be shown that the $n$th term of the current series \eqref{eq: final answer for current} is of order $J^{n+1}$ as $J\to 0$ and (for $n\ge2$) of order $1/ J^{n+1}$ as $|J|\to\infty$ ; this means that the series applies for both weak and strong coupling.

\emph{Steady state.} A basic question in quench problems is the existence of the steady-state limit of observable quantities, such as the current: $I_{\text{steady state}}(T_1,T_2, V) = \lim_{t \to\infty} I(T_1, \mu_1=0; T_2,\mu_2=-V; t)$.
We have shown that all orders of our series (in $J$ or in $1/J$) converge in the steady-state limit, and we have verified that the same series for the steady-state current is obtained by directly evaluating the current operator in the NESS.
Our results complement those of Doyon and Andrei  \cite{DoyonAndrei}, who showed that the Keldysh perturbation series for the current converges in time to all orders in $J$.

We proceed to investigate the steady-state current in the scaling regime, in which the external scales $T_1$, $T_2$, and $V$ are much smaller than the bandwidth. We express our answers.  We express our answers in terms of the usual $g\equiv \rho J = \frac{1}{2\pi}J$ \footnote{In our convention, $\rho=1/(2\pi)$ is the density of states per unit length.}.

First, we review what is expected.  In the regime of small $|g|$, the perturbative renormalizability of the Kondo model constrains the steady-state current to the form  $I_{\text{steady state}}(T_1,T_2,V) \to V \sum_{n=2, 0\le m \le n-2}^\infty\ a_{nm} g^n \ln^m \frac{2D}{M}$, where $M= \sqrt{\frac{1}{2}\left(T_1^2 + T_2^2\right) + V^2}$ and where the coefficients $a_{nm}$ depend only on the ratios $T_1/V$ and $T_2/V$.  (This is shown in a very general setting by Delamotte in Ref. \cite{Delamotte}; our choice of $V$ for the dimensionful prefactor and $2D/M$ for the argument of the log is one of convenience.)  Our calculation indeed produces a series of this form (see Supplemental Material \footnote{See Supplemental Material below for the series in $g$ and for the definition of some quantities that appear in the series in $1/g$.}), which we use as a check by comparing to known results in the universal antiferromagnetic regime.  To obtain a universal answer, we use the renormalization group (RG) scaling equation (or Callan-Symanzik equation) $\left[ D \frac{\partial}{\partial D} + \beta(g) \frac{\partial}{\partial g} + \gamma(g) \right] I_{\text{steady state}} = 0$.  The solution takes the form $I(T_1,T_2,V) =f_{universal}(T_1/T_K, T_2/T_K, V/T_K) e^{ -\int_0^g dg'\ \frac{\gamma(g')}{\beta(g')} }$, where the $g$-dependent scale factor goes to unity in the scaling limit ($g\to0^+$, $D\to\infty$ with $T_K$ fixed) because $\gamma(g)$, we find, starts at the same order in $g$ as $\beta(g)$.  (Such a scale factor has been seen before in the Kondo problem; see Ref. \cite{BarzykinAffleck}.)

Let us consider the differential conductance $G \equiv \partial I_{\text{steady state}}/\partial V$ in the scaling limit, focusing on the case of equal lead temperatures ($T_1=T_2$).  At the leading order, we obtain the standard result $G = G_0 \frac{3\pi^2 /16}{\ln^2 \frac{\sqrt{T^2 + V^2}}{T_K} }$ \cite{KaminskiNazarovGlazman}, where $G_0= 2e^2/h$ is the unitarity limit of conductance and $T_K = D e^{- 1/(2g)}$ is the Kondo temperature.  The next order corrections to $G$ and $T_K$ are affected by  our  cutoff scheme (see Refs. \cite{CulverAndrei_PRB1, Culver_thesis} for further discussion); however, the first correction beyond the leading order in the quantity $\Delta G(T,V) \equiv G(T,V) - G(T=0,V)$ agrees with the one-loop results of Ref. \cite{DoyonAndrei} after correcting some minor errors in Ref. \cite{DoyonAndrei}.

\emph{Universal strong ferromagnetic regime.}  Our approach reveals another universal regime of the Kondo model: strong ferromagnetic coupling ($g <0$, $|g|\gg 1$).  We note that there are proposals for mesoscopic realizations \cite{MitchellEtAl, BaruselliEtAl} of the weak ferromagnetic model (see also Ref. \cite{KuzmenkoEtAl}); it may be possible to realize the strong ferromagnetic model by modifying these proposals to use the charge Kondo effect \cite{MitchellPrivateCommunication}.

For strong coupling of either sign ($|g|\gg1)$, we obtain the following result at large bandwidth (presented with leading logs in the first row, subleading logs in the second row, etc.):
\begin{widetext}
        \small
        \begin{alignat}{3}
        I(T_1,T_2,V)=
            \frac{1}{\pi}V \Biggr\{ 1 - \frac{4}{9\pi^2} \biggr[ \frac{7}{g^2} &-\frac{16}{\pi^2 g^3} \ln\frac{2D}{M}  &&+\frac{64}{\pi^4 g^4}\ln^2 \frac{2D}{M}  &&-\frac{2048}{9\pi^6 g^5} \ln^3 \frac{2D}{M} \notag\\
            &-C_1\frac{16}{\pi^2 g^3}   &&+C_1\frac{128}{\pi^4 g^4}  \ln\frac{2D}{M}  &&+\left(4-12C_1 \right)\frac{512}{\pi^6 g^5}\ln^2\frac{2D}{M} \notag \\
            & &&+
            \left(3 C_2+ 6\pi \widetilde{C_1} - 22\pi^2
            \right)
            \frac{16}{9\pi^4 g^4} &&+
            \begin{pmatrix}
            32- 8 C_2
            +16 C_1 \\ - 12\pi \widetilde{C_1} + 11\pi^2 
            \end{pmatrix}
            \frac{64}{9\pi^6 g^5} \ln \frac{2D}{M} \notag\\
            & &&  &&+C_4 \frac{1}{g^5}+O\left(\frac{1}{g^6}\right)  \biggr] \Biggr\},
    \end{alignat}
    \normalsize
\end{widetext}
where the coefficients $C_1$, $\widetilde{C_1}$, $C_2$, and $C_4$ are functions of the ratios $T_1/V$ and $T_2/V$ (another function $C_3$ appears in the series for small $g$; see Supplemental Material).  In the case of equal lead temperatures ($T_1=T_2$), we find that the RG scaling equation holds with $\beta(g) = -\frac{8}{3\pi^2}\left[ 1 + \frac{32}{9 \pi^2 g}  +O\left(1/g^2\right) \right]$ and $\gamma(g) = \frac{256}{27\pi^4 g^3}\left[ 1 + \frac{56}{9\pi^2 g} +O\left( 1/g^2\right) \right]$, and thus the following Kondo temperature $T_K^{(F)} = D e^{-\int^g dg'\ \frac{1}{\beta(g')}}$ for this regime \footnote{It is interesting to note that the same Kondo temperature can be read off from the Bethe ansatz solution of the equilibrium problem (with the regularization scheme and notation used in Ref. \cite{AndreiFuruyaLowenstein}). The absence of a $\ln |g|$ term in the exponent of the antiferromagnetic $T_K$  is well-known in the Bethe ansatz approach \cite{AndreiFuruyaLowenstein}.}:
\begin{equation}
    T_K^{(\text{F})} \equiv D e^{\frac{3\pi^2}{8}g  - \frac{4}{3} \ln |g|}.\label{eq: TK strong ferro}
\end{equation}
Notice that we can take the scaling limit $D\to\infty$, $g\to-\infty$ with $T_K^{(\text{F})}$ held fixed, indicating that the strong ferromagnetic regime is universal.

\begin{figure}[htp]
    \centering
    \includegraphics[width=\linewidth]{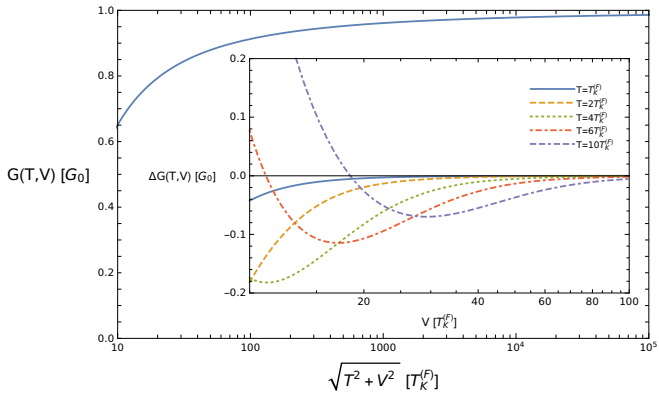}
    \caption{The universal conductance $G \equiv \pd I_{\text{steady state}}/\pd V$ in the strong ferromagnetic regime at leading log approximation.  In contrast to the antiferromagnetic case in which $G$ is known to reach the unitarity limit $G_0 \equiv 2e^2/h$ at $T=V=0$ \cite{GlazmanRaikh}, here the unitarity limit is reached asymptotically at \emph{large} voltage or temperature.  As the external scale is lowered to $T_K^{(\text{F})}$ and below, our series in $1/g$ breaks down and another method is needed.  Inset: The first correction beyond leading logs in the quantity $\Delta G \equiv G(T,V) - G(T = 0,V)$ for $V\gg T_K^{(\text{F})}$, with various values of $T$.}\label{fig: strong ferro combined}
\end{figure}

Resumming the leading logs of the current series, we find that the conductance approaches the unitarity limit asymptotically at \emph{high} voltage or temperature (Fig. \ref{fig: strong ferro combined}):
\begin{equation}
    G(T,V) = G_0 \left( 1 - \frac{3\pi^2}{16 \ln\frac{ \sqrt{T^2 +V^2}}{T_K^{(\text{F})}} } + \dots\right).\label{eq: G leading log strong ferro}
\end{equation}
In analogy to the antiferromagnetic case, we expect that the coefficient $-\frac{4}{3}$ of $\ln |g|$ in Eq. \eqref{eq: TK strong ferro} is affected by our cutoff scheme; however, any change of this coefficient would only affect higher-order corrections to Eq. \eqref{eq: G leading log strong ferro}.  We expect that in the first correction, the difference $\Delta G$ is reliable (see the inset of Fig. \ref{fig: strong ferro combined}), as this quantity was verified in the antiferromagnetic case.

\emph{Models with charge fluctuations.} We briefly summarize the results of applying our method to the interacting resonant level model (IRL), $H_{\text{IRL}} = H_{\text{leads}} + \epsilon d^\dagger d +  \text{Re}\left\{ 2\sqrt{\Delta}  [ \psi_1^\dagger(0) + \psi_2^\dagger(0) ] d \right\}  + U [ \psi_1^\dagger(0) \psi_1(0)+ \psi_2^\dagger(0) \psi_2(0)]d^\dagger d  $, and the Anderson impurity model (AIM), $H_{\text{AIM}} = H_{\text{leads}} + \epsilon d_a^\dagger d_a +  \text{Re}\left\{ 2\sqrt{\Delta}  [ \psi_{1a}^\dagger(0) + \psi_{2a}^\dagger(0) ] d_a \right\}  + U d_\uparrow ^\dagger d_\uparrow d_\downarrow ^\dagger d_\downarrow$ [where $H_{\text{leads}}$ is the same kinetic term as in Eq. \eqref{eq: two-lead Kondo}, omitting the spin index in the IRL case].  Details of our calculations are reported in Refs. \cite{CulverAndrei_PRB2, Culver_thesis}.

In the IRL, we find the exact time-evolving wavefunction after a quench that switches on $\epsilon$, $\Delta$, and $U$ at $t=0$.  We evaluate the steady-state occupancy $\langle d^\dagger d \rangle$ to leading order in $U$, and show that it is universal with the standard scale $T_K^{(\text{IRL})} \sim  D \left(\frac{\Delta}{D}\right)^{1/(1+  U/\pi)}$.  In the equilibrium limit (i.e., zero temperature and voltage), our result agrees with the Bethe ansatz result from the literature \cite{Ponomarenko} (see also \cite{RylandsAndrei} and \cite{CamachoSchmitteckertCarr}).  Out of equilibrium, we find that that the series in $U$ breaks down at a very large voltage $V_0 \sim T_K^{(\text{IRL})} e^{2/(\rho U)}$ (where $\rho=1/2\pi$ is the density of states per unit length).  This scale $V_0$ could also be significant in the lattice model if it lies in the universal regime, i.e., if $V_0 \ll D_{\text{lattice}}$.

In the AIM, we calculate the NESS wavefunction directly either for small $U$ or infinite $U$.  The small $U$ expansion of the steady-state current is found to be $I_{\text{s.s.}} = I_{\text{s.s.}}^{(0)} + I_{\text{s.s.}}^{(1)} +\dots$, where $I_{\text{s.s.}}^{(0)}$ is the standard resonant level current and
\begin{multline}
    I_{\text{s.s.}}^{(1)} = \frac{U}{8 \Delta^2} \int_{-D}^D \frac{dk_1}{2\pi}\frac{dk_2}{2\pi}\ \left[\fermifn_1(k_1) + \fermifn_2(k_1) \right]\\
    \times \left[\fermifn_1(k_2) - \fermifn_2(k_2) \right] |\mathcal{T}(k_1)|^2 |\mathcal{T}(k_2)|^2 \text{Re}\left[\mathcal{T}(k_2) \right],
\end{multline}
where $\mathcal{T}(k) = 2\Delta/(k - \epsilon + i \Delta)$ is the single-electron $\mathcal{T}$ matrix of this model.  We verify $I_{\text{s.s.}}^{(1)}$ by Keldysh perturbation theory.  For infinite $U$, we find an expansion for the steady-state current in powers of $\Delta$, with the standard scaling invariant $\epsilon_d \equiv \epsilon + \frac{\Delta}{\pi} \ln \frac{D}{\Delta}$ \cite{Haldane_PRL, Hewson}.

\emph{Discussion.} We provided an exact, explicit solution for the time-evolving wavefunction in the nonequilibrium Kondo model.  We obtained a series expression for the current which can be expanded either for weak coupling or strong coupling, and used it to explore another universal regime.  It still should be checked that this regime exists in the lattice model.  To see the predicted rise of the conductance towards the unitarity limit, one would need a hierarchy of scales $T_K^{(\text{F})} \ll V \ll E_{\text{max}}$ or $T_K^{(\text{F})} \ll T \ll E_{\text{max}}$, where $E_{\text{max}}$ is the energy scale beyond which the Kondo model is no longer an accurate description of the system.

\begin{figure}[htp]
    \centering
    \includegraphics[width=\linewidth]{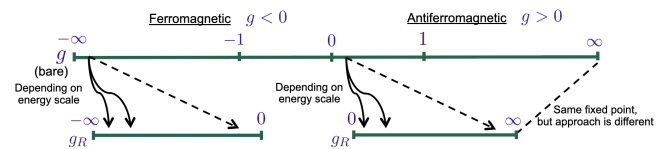}
    \caption[Kondo scaling picture]{Kondo scaling picture.  The two universal regimes are weak antiferromagnetic \emph{bare} coupling ($0 < g \ll 1$, $T_K = D e^{-1/(2g)}$) and strong ferromagnetic \emph{bare} coupling ($g<0$, $|g| \gg1$, $T_K^{(\text{F})} = D e^{- 3\pi^2 |g| /8}$).  The former has been much studied, and the latter is predicted by our calculations.  In either case, the running coupling $g_R$ is close to the bare coupling if the system is probed at a high energy scale (high relative to $T_K$ or $T_K^{(F)}$, though always small compared to the bandwidth), and moves away from the bare coupling as the energy scale is reduced.}
    \label{fig: Kondo scaling}%
\end{figure}

We have the following picture of the RG flow in the strong ferromagnetic regime (Fig. \ref{fig: Kondo scaling}).
Starting at the unstable fixed point $g_R=-\infty$, the running coupling $g_R$ becomes \emph{smaller} in magnitude according to $g_R = - \frac{8}{3\pi^2} \ln \frac{T}{T_K^{(\text{F})}}$ (at leading order).  As $T$ approaches $T_K^{(\text{F})}$ from above, $|g_R|$ becomes too small for our calculation to be valid.  We expect, though, that $g_R$ continues to flow to the stable fixed point $g_R=0^-$ without any other fixed points in between (much as the corresponding antiferromagnetic flow from $g_R = 0^+$ to $g_R=\infty$).  The ground state of the system would flow from a triplet at high energy, with entropy $\ln 3$, to a free spin at low energy, with entropy $\ln 2$.

It would be interesting to see if our method for calculating local quenches and nonequilibrium steady states can be useful in a wider class of problems.  We note that the usual signatures of integrability in the Kondo model, such as the Yang-Baxter equation, do not appear in any obvious way in our calculations.    

\emph{Ackowledgements.} We are grateful to Chung-Hou Chung, Piers Coleman, Garry Goldstein, Yashar Komijani, Yigal Meir, Andrew Mitchell, Achim Rosch, and Hubert Saleur for helpful discussions, and to Sung Po Chao for comments on an earlier draft.  We have benefited from working on related problems with Huijie Guan, Paata Kakashvili, Christopher Munson, and Roshan Tourani.  A.B.C. acknowledges support from the Samuel Marateck Fellowship in Quantum Field Theory Physics and the Excellence Fellowship (both from Rutgers University).  This material is based upon work supported by the National Science Foundation under Grant No. 1410583.

\bibliographystyle{apsrev4-2}
\bibliography{references}

\begin{widetext}

\section{Supplemental Material for ``Many-body wavefunctions for quantum impurities out of equilibrium''}

The content of this Supplemental Material is derived in detail in Refs. \cite{CulverAndrei_PRB1, Culver_thesis}.

The steady state current expanded in powers of $g$ is found to be:
\small
    \begin{alignat}{3}
            I(T_1,T_2,V)=
            \frac{3\pi}{4} V \Biggr\{ g^2 &+4 g^3 \ln\frac{2D}{M}  &&+12 g^4 \ln^2 \frac{2D}{M}  &&+32 g^5 \ln^3 \frac{2D}{M} \notag\\
            &+ C_1\left( \frac{T_1}{V},\frac{T_2}{V} \right) g^3   &&+6 C_1\left( \frac{T_1}{V},\frac{T_2}{V} \right) g^4\ln\frac{2D}{M}  &&+\left[24C_1\left( \frac{T_1}{V},\frac{T_2}{V} \right) -32 \right] g^5 \ln^2\frac{2D}{M} \notag \\
            & &&+ C_2\left( \frac{T_1}{V},\frac{T_2}{V} \right) g^4  &&-
            \begin{pmatrix}
            16 C_1\left( \frac{T_1}{V},\frac{T_2}{V} \right) - 8C_2\left( \frac{T_1}{V},\frac{T_2}{V} \right) \\ +64 + 3\pi^2
            \end{pmatrix}
            g^5\ln\frac{2D}{M}\notag\\
            & &&  &&+C_3\left(  \frac{T_1}{V},\frac{T_2}{V} \right)\ g^5  +O(g^6)   \Biggr\} ,\label{eq: I(T1,T2,V)}
        \end{alignat}
        \normalsize
where the $C$ functions (some of which also appear in the $1/g$ series in the main text) are defined as follows.  Replace $\left(\frac{T_1}{\sqrt{2}} ,\frac{T_2}{\sqrt{2}},V\right)$ by spherical coordinates $(M,\theta,\phi)$:
\beq
        M = \sqrt{\frac{1}{2}\left( T_1^2 + T_2^2 \right) +V^2 },\ \theta= \arctan \frac{\frac{1}{2}(\sqrt{T_1^2 + T_2^2)}}{V},\ \phi = \arctan \frac{T_2}{T_1},
\eeq
then define:
\bseq
\begin{align}
    f(\theta, \phi; v) &= \frac{\sqrt{2} \pi \sin \theta \cos \phi\ v e^{-i (\cos \theta) v} }{\sinh(2^{3/2} \pi \sin\theta \cos \phi\ v  )} + \frac{\sqrt{2} \pi \sin \theta \sin \phi\ v e^{i (\cos \theta) v} }{\sinh(2^{3/2} \pi \sin\theta \sin \phi\ v)  },\label{eq: fn f Kondo}\\
    h(\theta,\phi; v) &=  \frac{1}{i} \left( \frac{\sqrt{2} \pi \tan \theta \sin \phi\ e^{i (\cos \theta) v} }{\sinh(2^{3/2} \pi \sin\theta \sin \phi\ v  )} - \frac{\sqrt{2} \pi \tan \theta \cos \phi\ e^{-i (\cos \theta) v} }{\sinh(2^{3/2} \pi \sin\theta \cos \phi\ v)  } \right).\label{eq: fn h Kondo}
\end{align}
\eseq
Then:
\bseq
\begin{align}
    C_1\left(\theta,\phi\right) &= 4\ \text{Re}\left\{ \gamma - \int_0^\infty du\ \ln u \frac{\partial}{\partial u} \left[ f\left(\theta,\phi;u\right) h\left(\theta,\phi;-u\right) \right] \right\},\label{eq: C1}\\
    \widetilde{C}_1\left(\theta,\phi\right) &= 4\ \text{Im}\left\{ \gamma - \int_0^\infty du\ \ln u \frac{\partial}{\partial u} \left[ f\left(\theta,\phi;u\right) h\left(\theta,\phi;-u\right) \right] \right\},\label{eq: C1Tilde}\\
    C_2\left(\theta,\phi\right) &= \text{Re}\biggr\{ 6\gamma C_1\left(\theta,\phi\right) - 12\gamma^2 + \frac{7}{12}\pi^2- 4\int_0^\infty du\ \ln^2 u \frac{\partial}{\partial u}\left[ f\left(\theta,\phi;u\right) h\left(\theta,\phi;-u \right) \right] \notag\\
    &+ 8 \int_0^\infty du_1 du_2\ \ln u_1 \ln u_2 \frac{\partial }{\partial u_1}\frac{\partial }{\partial u_2} \left[ f\left(\theta,\phi,u_1 \right) f\left(\theta,\phi,u_2\right) h\left(\theta,\phi;-u_1 -u_2 \right) \right]\notag \\
    & +8 \int_0^\infty du_1 du_2\ \frac{1}{u_2} \ln \frac{u_1 +  u_2}{u_1} \frac{\partial}{\partial u_1} \left[ f\left(\theta,\phi,u_1 + u_2\right) f\left(\theta,\phi,-u_1\right) h\left(\theta,\phi;-u_2\right) \right] \biggr\},\label{eq: C2}
\end{align}
\eseq
where $\gamma$ is the Euler constant.  We omit very lengthy expressions for $C_3$ and $C_4$ (they are again integrals over $f$ and $h$, now including triple integrals).

\end{widetext}

\end{document}